\documentclass{mem}
\usepackage{natbib}
\usepackage{balance}
\usepackage{graphicx}
\usepackage[a4paper]{hyperref}
\idline{75}{282}
\begin{document}
\def\teff{$T\rm_{eff }$}
\def\kms{$\mathrm {km s}^{-1}$}

\title{
The Composition of RR Lyrae Stars: Start-line for the AGB
}

   \subtitle{}

\author{
George Wallerstein 
\and Wenjin Huang 
          }


\institute{
Deparment of Astronomy, University of Washington\\
Box 351580, Seattle, WA 98195\\
wall@astro.washington.edu; hwenjin@astro.washington.edu
}

\authorrunning{Wallerstein \& Huang}

\titlerunning{RR Lyrae Stars}

\abstract{
This paper sumarizes research  on abundances in RR Lyrae stars that
one of us (GW) has been engaged in with various astronomers.  In addition
we report on preliminary analysis of the abundances of C, Si, S and Fe in
24 RR Lyrae stars.  Our model atmosphere analysis, including NLTE effects,
are based on the spectra of resolving power 30,000 obtained at the Apache
Poing Observatory. 
}
\maketitle{}

\section{Introduction}
This paper reports on research by Sergei M. Andrievsky, Valentin V. Kovtyukh,
Marcio Catelan, Dana Casetti-Dinescu, and Gisella Clementini as well as ourselves.

The RR Lyrae stars are of great value in studies of the structure of our
Galaxy and the history of its composition as modified by stellar
nucleosynthesis over the ages. The fact that they show a narrow range of
luminosity from about Mv=0.4 to 0.8, as derived from their membership in
globular clusters, allows their distances to be derived and their orbits
calculated from their proper motions and radial velocities. Their effective
temperatures vary from approximately 6000 to 7000 K during their pulsation.
Their spectra often show lines of many elements without excessive blending.

The chemical compostion of RR Lyrae stars is a combination of their original
composition when they were main sequence stars near Mv=4.5 with important changes
induced by nuclear reactions whose products may be convected to the stellar
surface. The first mixing event to affect their atmospheres is the deepening
of convection as the star leaves the subgiant branch and begins its evolution
up the almost vertical giant branch \citep{hoy55}.
CNO-processing converts some carbon into nitrogen and reduces the 12C/13C
ratio from its initial value to about 20. As the star's evolution slows
down at the red giant clump additional mixing further reduces the carbon
and lowers the 12C/13C ratio to about 4-8 \citep{gil91,cha94,cha98}. These
changes have been seen in many globular clusters.

At the tip of the red giant branch important events take place. The triple-alpha
reaction starts in the degenerate core producung additional carbon.  At first the
tempature rises exponentially because the pressure and density do not respond
until the degeneracy is removed by the temperature rise and the reaction rate
depends approximately on the temperature raised to the 30th power.  This sets up
a grossly superadiabatic temperature gradient and initiates violent convection
\citep{moc09}.  Calculations indicate that the products of helium burning
do not reach the surface but observations of the carbon abundance in RR Lyrae
stars are useful to confirm the accuracy of the calculations and to describe
the initial conditions for stars that will eventually evolve up the AGB, perhaps
to become carbon stars.  In addition mass-loss on the red giant branch plus
additional mass-loss at the time of the helium-flash reduces the stellar mass
from its original value of about 0.8 Msun to about 0.55 Msun as derived from
multi-periodic RR Lyrae stars. Such mass-loss may give us a glimpse into the
interior of what had once been a red giant.

Our first project to determine the composition of RR Lyrae stars was suggested
by M. Catelan who noted that a few RR Lyrae and related stars showed kinematics
similar to that of Omega Cen. The first is VY Ser, a genuine RR Lyrae star
of period 0.714 days, while 2 more, V716 Oph and XX Vir have periods slightly
 longer than 1.0 days and are usually called short-period type II cepheids, 
but might be referred to as long-period RR Lyrae stars. Dana Casetti-Dinescu
confirmed that their galactic orbits do indeed relate them to Omega Cen. The
period of VY Ser is typical of RR Lyrae stars in Omega Cen, but only 7
short-period cepheids are known in Omega Cen \citep{cle01}.  In fact it 
is remarkable that astronomers have defined the break between RR Lyraes and
short period cepheid to equal to the period of rotation of the earth.   
Our observations consisted of 4 echelle spectra of VY Ser, 5 of V716 Oph,
and 2 of XX Vir that have been analysed by Andrievsky and Kovtyukh. All 3 stars
show [Fe/H] close to $-1.6$ which is typical of Omega Cen but cannot be used
to reveal the strong gradient of [s/Fe] when plotted against [Fe/H] in
Omega Cen \citep{van94, nor95}.

Our second group of targets was RR Lyrae stars with period greater than 0.75 days.
Such stars are very rare in globular clusters with the conspicuous exception of 
the two unusual clusters, NGC 6388 and 6441 \citep{pri01, pri02}.  These
relatively metal rich clusters with [Fe/H] near $-0.8$ have numerous long-period  
 RR Lyrae stars. In the general field such stars are rare but some are known and
their light curves have been obtained by \citet{sch02}.  Only a few are
 sufficiently bright for highres spectroscopy with the 3.5-m telescope of the
Apache Point Observatory. They are listed in \citet{wal09}.  [Fe/H] values were
found for 4 stars but their metallicities ranged from $-1.8$ to $+0.2$.
Hence our analyses of these stars failed to relate them to NGC 6388 and 6441.  
KP Cyg remains an almost unique RR Lyrae star with a period of 0.856 days and
[Fe/H] $= +0.2$ on the basis of 5 spectra well distributed in phase. It had
 already been recognized as having delta-S $= 0$ by \citet{pre59}. Our spectra show
a typical velocty curve for an RR Lyrae star with H-alpha emission at phases
0.43, 0.72, and 0.80.

The third aspect of our RR Lyrae analyses was the recognition that the relatively
metal-rich stars show an apparant excess of carbon \citep{wal09}).  The carbon
abundance was derived from the lines around 7115\AA~of multiplets 108 and 109
of \citet{wie98}.  These lines are  sufficiently weak to be
found only in stars with [Fe/H] $> -1.0$. For metal-poor 
stars stronger lines must be employed which suffer from NLTE \citep{fab06}.
Fabbian has told us (private communication) that the lines around 7115\AA~should
not suffer from significant NLTE effects. The best strong lines are one at
8335\AA~and from multiplet 62 beyond 9000\AA.  That spectral region contains many
moderately strong atmospheric H$_{\rm 2}$O lines. 

\section{Observation and Data Reduction}
New Observations were obtained with the echelle spectrograph on the Apache Point
3.5-m telescope in March, November and December, 2009.  The resolving power is about
30,000 and the signal-to-noise ratio of the reduced spectra is usually 100-150.
The cold temperatures and altitude of 9200 ft permit very effective division by
the spectrum of a hot rapidly rotating star so as to  cancel out the atmospheric
H$_{\rm 2}$O absorption lines, especially beyond 8900\AA.
We have used the H$\gamma$ line to establish Teff and the known mass
 and luminosity of the RR Lyrae stars for the derivation of the surface gravity
(without introducing the perturbation of $\log g$ by the acceleration induced by  
pulsation). In addition to deriving the iron abundance from Fe II lines we have
compared the C I lines with lines of S I whose excitation and ionization potentials
are similar to those of C I. We added Si II to the data base as a check on the 
alpha element excess usually seen in metal-poor stars. The observed lines and their
atomic properties are shown in Table 1.  The abundances were derived using the
updated version of the line synthesis code, MOOG, with the Kurucz atmosphere models
\citep{cas04}.  In Table 2 we show the derived abundances 
of C, S, Si, and Fe. We also show the Delta-S value from \citet{pre59} or other
sources such as \citet{lay94}.  The correlation of the Fe abundance with delta-S
is very good.

\begin{table*}
\caption{Observed lines}
\label{linelist}
\begin{center}
\begin{tabular}{crrr}
\hline
\\
Wavelength &         & $\chi$ &           \\
(\AA)      & Ions &  (eV)  & $\log gf$ \\
\hline
\\
4932.05 &   6.00 &   7.68 &  $-$1.658  \\
5052.17 &   6.00 &   7.68 &  $-$1.303  \\
5380.34 &   6.00 &   7.68 &  $-$1.616  \\
7111.47 &   6.00 &   8.63 &  $-$1.085  \\
7113.18 &   6.00 &   8.64 &  $-$0.773  \\
7115.17 &   6.00 &   8.63 &  $-$0.824  \\
7116.99 &   6.00 &   8.64 &  $-$0.907  \\
7119.66 &   6.00 &   8.63 &  $-$1.148  \\
8335.15 &   6.00 &   7.68 &  $-$0.437  \\
9061.44 &   6.00 &   7.47 &  $-$0.347  \\
9062.49 &   6.00 &   7.47 &  $-$0.455  \\
9078.29 &   6.00 &   7.47 &  $-$0.581  \\
9088.51 &   6.00 &   7.47 &  $-$0.430  \\
9094.83 &   6.00 &   7.48 &   0.151  \\
9111.81 &   6.00 &   7.48 &  $-$0.297  \\
9405.73 &   6.00 &   7.68 &   0.286  \\
3853.66 &  14.01 &   6.86 &  $-$1.341  \\
5957.56 &  14.01 &  10.07 &  $-$0.225  \\
5978.93 &  14.01 &  10.07 &   0.084  \\
6347.09 &  14.01 &   8.12 &   0.149  \\
6371.36 &  14.01 &   8.12 &  $-$0.082  \\
8694.70 &  16.00 &   7.87 &   0.154  \\
9212.91 &  16.00 &   6.52 &   0.430  \\
9237.54 &  16.00 &   6.52 &   0.030  \\
\hline
\end{tabular}
\end{center}
\end{table*}

\begin{table*}
\caption{Abundances of Observed RR Lyrae Stars}
\label{abu}
\begin{center}
\begin{tabular}{lccrrrrrrrr}
\hline
\\
Star    & $P$(days) & $T_{\rm eff}$/$\log g$/$V_{\rm t}$ & $\Delta S$& [Fe/H]   &   [C/Fe]  &  [S/Fe]  & [Si/Fe] &    [C/S] &  [C/Si] &    Note \\
\hline
\\
V445OPH & 0.397 & 6500/2.5/2.2 &   1    &    0.24  &  $-$0.39  & $-0.22$  & $-0.07$ &  $-0.17$ &  $-0.32$ &    1,3  \\
RRGEM   & 0.397 & 6750/2.5/3.1 &   3    &    0.01  &  $-$0.39  & $-0.44$  & $-0.20$ &  $ 0.05$ &  $-0.19$ &    1,3  \\
SWAND   & 0.442 & 6500/2.5/4.0 &   0    & $-$0.16  &  $-$0.39  & $-0.56$  & $-0.06$ &  $ 0.17$ &  $-0.33$ &     3   \\
DXDEL   & 0.473 & 6500/2.5/3.1 &   2    & $-$0.21  &  $-$0.28  & $-0.20$  & $-0.18$ &  $-0.08$ &  $-0.10$ &    1,3  \\
ARPER   & 0.426 & 6500/2.5/4.0 &   0    & $-$0.32  &  $-$0.32  & $-0.51$  & $ 0.01$ &  $ 0.19$ &  $-0.33$ &     3   \\
KXLYR   & 0.441 & 7000/3.0/3.1 &   0    & $-$0.57  &  $-$0.22  & $-0.27$  & $ 0.35$ &  $ 0.05$ &  $-0.57$ &    2,3  \\
XZDRA   & 0.476 & 6500/2.5/3.0 &   3    & $-$0.75  &     0.06  & $ 0.09$  & $ 0.33$ &  $-0.03$ &  $-0.30$ &    2,3  \\
UUVIR   & 0.476 & 6250/2.5/3.2 &   2    & $-$0.90  &  $-$0.42  & $ 0.11$  & $ 0.52$ &  $-0.53$ &  $-0.94$ &         \\
BHPEG   & 0.641 & 6500/2.5/2.2 &   6    & $-$1.17  &  $-$0.53  & $ 0.10$  & $ 0.40$ &  $-0.63$ &  $-0.93$ &         \\
WCVN    & 0.552 & 6250/2.5/3.0 &   7    & $-$1.22  &  $-$0.46  & $-0.01$  & $ 0.42$ &  $-0.45$ &  $-0.88$ &         \\
VZHER   & 0.440 & 6250/2.5/2.6 &   4    & $-$1.30  &  $-$0.23  & $ 0.21$  & $ 0.60$ &  $-0.44$ &  $-0.83$ &         \\
RVUMA   & 0.468 & 6500/2.5/2.3 & 3.5    & $-$1.31  &  $-$0.05  & $ 0.22$  & $ 0.58$ &  $-0.27$ &  $-0.63$ &         \\
RRLEO   & 0.452 & 6500/2.5/2.8 &   5    & $-$1.39  &  $-$0.04  & $ 0.22$  & $ 0.61$ &  $-0.26$ &  $-0.65$ &         \\
TTLYN   & 0.597 & 6500/2.5/3.4 &   7    & $-$1.41  &  $-$0.98  & $-0.24$  & $ 0.12$ &  $-0.74$ &  $-1.10$ &         \\
RRLyr   & 0.567 & 6500/2.5/4.0 &   6    & $-$1.44  &  $-$0.91  & $-0.03$  & $ 0.14$ &  $-0.88$ &  $-1.05$ &         \\
TUUMA   & 0.558 & 6500/2.5/3.4 &   6    & $-$1.46  &  $-$0.32  & $-0.04$  & $ 0.69$ &  $-0.28$ &  $-1.01$ &         \\
VXHER   & 0.455 & 6000/2.5/2.4 &   5    & $-$1.48  &  $-$0.34  & $ 0.16$  & $ 0.59$ &  $-0.50$ &  $-0.93$ &         \\
DHPEG   & 0.256 & 6500/2.5/3.0 &   0    & $-$1.53  &  $-$0.12  & $ 0.23$  & $ 0.75$ &  $-0.35$ &  $-0.87$ &         \\
RRCET   & 0.553 & 6500/2.5/3.7 &   5    & $-$1.61  &     0.15  & $ 0.09$  & $ 0.85$ &  $ 0.06$ &  $-0.70$ &         \\
STBOO   & 0.622 & 6250/2.5/4.0 &   9    & $-$1.77  &  $-$0.50  & $-0.15$  & $ 0.46$ &  $-0.35$ &  $-0.96$ &         \\
SVERI   & 0.714 & 6500/2.5/3.0 &   9    & $-$1.94  &  $-$0.44  & $-0.15$  & $ 0.12$ &  $-0.29$ &  $-0.56$ &         \\
RUPSC   & 0.390 & 6500/2.5/3.5 &   7    & $-$2.04  &  $-$0.62  & $-0.12$  & $ 0.51$ &  $-0.50$ &  $-1.13$ &         \\
RZCEP   & 0.511 & 6500/2.5/3.0 &   5    & $-$2.10  &  $-$0.41  & $ 0.01$  & $ 0.57$ &  $-0.42$ &  $-0.98$ &         \\
XARI    & 0.651 & 6250/2.5/3.8 &  10    & $-$2.68  & $<-$0.46  & $ 0.01$  & $ 0.55$ & $<-0.47$ & $<-1.01$ &         \\
\hline
\end{tabular}
{\footnotesize
 \begin{enumerate}
      \item Si II 5957, 5978 only
      \item Si II 3853, 5957, 5978 only
      \item C I 7100 lines only
   \end{enumerate}
}
\end{center}
\end{table*}

\section{Results}
In Fig.\ 1(a) we show [Si/Fe] plotted against [Fe/H]. The pattern is similar to that
which is usually found for metal-poor stars.  In Fig.\ 1(b) the behavior of
[S/Fe] is qualitatively similar to that of [Si/Fe] but is displaced downward by
about 0.4 dex.  We do not fully understand the low [S/Fe] values because we have
already applied the NLTE corrections of \citet{tak05} to the derived sulfur abundance.
Fig.\ 1(c) shows [C/Fe] vs.\  [Fe/H] for these stars.  The scatter is unfortunately large
possibly because we have ignored the pulsation effect on $\log g$.  However, in Fig. 1(c),
it is evident that relatively high carbon abundances appear in the high [Fe/H] region.
Note that we also used the NLTE corrections of \citet{fab06} in our carbon abundance
calculations.  Since both the ionization potential and the excitation potentials of the
lines are similar, the effects of $\log g$ and possible NLTE uncertainties should be
diminished in the abundance ratios of [C/S] and [C/Si].  In Fig.\ 1(d), a clear
trend of increasing [C/S] is present.  The same trend is seen even more clearly
in the plot of [C/Si] vs.\ [Fe/H] (Fig.\ 1(e)).  The Si abundances come from the
ionized lines of similar excitation to the C I lines (see Table 1). 
 
\begin{figure*}[t!]
\resizebox{\hsize}{!}{\includegraphics[clip=true]{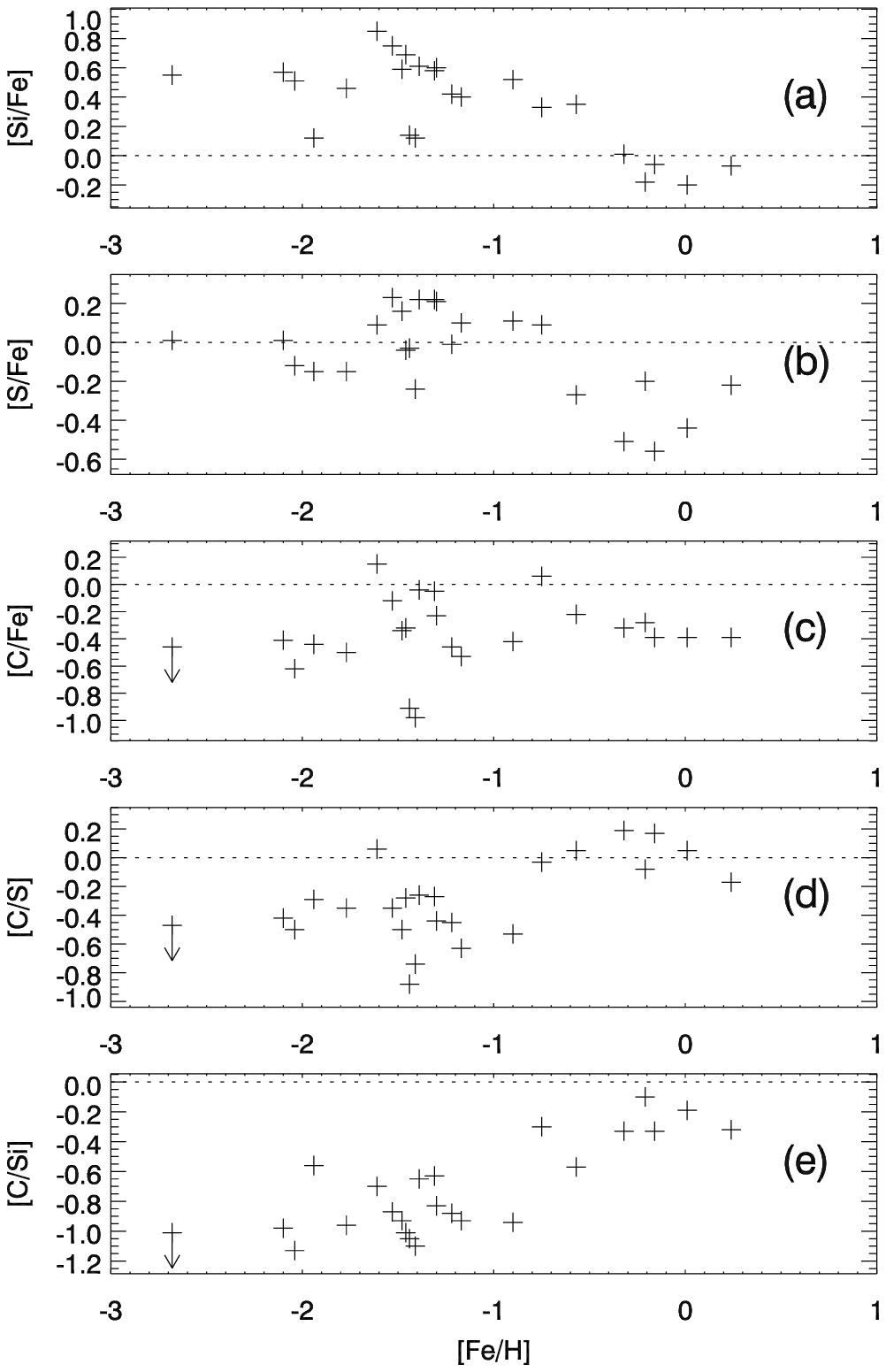}}
\caption{
\footnotesize
Plots of abundance ratios versus [Fe/H] based on Table~2:
(a) Our derived [Si/Fe] ratios based on ionized lines of both elements;
(b) Our derived [S/Fe] ratios based on S I and Fe II lines; (c) Our
derived [C/Fe] ratios based on C I and Fe II lines; (d) Our derived
[C/S] ratios based on C I and S I lines; (e) Our derived [C/Si] ratios
based on C I and Si II lines. }
\label{abn}
\end{figure*}

\begin{acknowledgements}
We are grateful to Damian Fabbian and Elisabetta Caffau for helpful
advice regarding NLTE in C I and S I, and to Sergei Andrievsky for
helpful suggestions.  This research was supported by the Kenilworth
Fund of the New York Community Trust.

\end{acknowledgements}

\bibliographystyle{aa}

\end{document}